\documentclass{article}

\usepackage[final,nonatbib]{neurips_2025_ml4ps}

\usepackage[utf8]{inputenc} 
\usepackage[T1]{fontenc}    

\usepackage{url}            
\usepackage{booktabs}       
\usepackage{amsfonts}       
\usepackage{nicefrac}       
\usepackage{microtype}      
\usepackage{xcolor}         
\usepackage{graphicx}  

\usepackage{subcaption}

\usepackage[backend=biber, citestyle=numeric, bibstyle=numeric]{biblatex}
\usepackage{xcolor} 
\usepackage[
    colorlinks=true,   
    linkcolor=blue,    
    citecolor=blue,    
    urlcolor=blue     
]{hyperref}
    
\usepackage[title]{appendix}
\usepackage{amsmath}

\title{Symbolic Regression Is All You Need: From Simulations to Scaling Laws in Binary Neutron Star Mergers}
\author{
Phelipe Darc*\\
Artificial Intelligence for Physics Laboratory (Lab-IA)\\
and Centro Brasileiro de Pesquisas F\'isicas (CBPF)\\
Rua Xavier Sigaud, 150, Urca, Rio de Janeiro, Brazil. \\
\And
Clecio R.~Bom\\
Artificial Intelligence for Physics Laboratory (Lab-IA)\\
and Centro Brasileiro de Pesquisas F\'isicas (CBPF)\\
Rua Xavier Sigaud, 150, Urca, Rio de Janeiro, Brazil. \\
\And
Charles Kilpatrick\\
Center for Interdisciplinary Exploration and Research in Astro-
physics (CIERA)\\
and Department of Physics and Astronomy\\
North-western University, Evanston, IL 60208, USA.\\
\And
Bernardo M. O. Fraga\\
Artificial Intelligence for Physics Laboratory (Lab-IA)\\
and Centro Brasileiro de Pesquisas F\'isicas (CBPF)\\
Rua Xavier Sigaud, 150, Urca, Rio de Janeiro, Brazil. \\
\And
Gabriel S. M. Teixeira\\
Artificial Intelligence for Physics Laboratory (Lab-IA)\\
and Centro Brasileiro de Pesquisas F\'isicas (CBPF)\\
Rua Xavier Sigaud, 150, Urca, Rio de Janeiro, Brazil. \\
}

\begin{document}

\maketitle

\begin{abstract}

Gravitational wave sources with electromagnetic counterparts have highlighted the need for predictive, interpretable models linking the parameters of compact binary systems to post-merger remnants and mass outflows. In this work, we explore AI-driven symbolic regression (SR) frameworks to derive updated analytical relations for disk ejecta mass in binary neutron star mergers, trained on state-of-the-art numerical relativity simulations. Our method reveals a set of compact equations that outperform existing fitting formulae across multiple statistical metrics while remaining physically interpretable. Notably, SR also enables alternative predictor sets (e.g., $\{M_1,M_2,\tilde{\Lambda}\}$) that match or exceed the accuracy of models relying solely on compactness of the lightest neutron star ($C_1$), enabling new parameter constraints from electromagnetic observations. Unlike traditional black-box machine learning models, these closed-form expressions generalize robustly to regions of the parameter space not represented in the training data, offering a physics-informed tool for multimessenger observations and constraints on the neutron star equation of state.

\end{abstract}

\section{Introduction}


The coalescence of binary neutron stars (BNS) and black hole–neutron star (BHNS) systems are prime sources of gravitational waves (GWs) observed by the Advanced LIGO and Virgo detectors \cite{Aasi_2015, Acernese2014}. These events are often accompanied by electromagnetic (EM) counterparts, including kilonovae. Kilonovae are powered by the radioactive decay of \emph{r}-process elements synthesized in neutron-rich ejecta, whose brightness and evolution depend on their total dynamical ejecta mass ($M_{\mathrm{dyn}}$), the mass remaining bound in the remnant accretion disk ($M_{\mathrm{disk}}$), and outflows from the disk.

Kilonova observations provide a unique window into the physical conditions of the merger remnant, including constraints on the mass, composition, and geometry of the ejected material.  Accurately modeling the ejecta properties is essential for applying EM signals to infer key source parameters, including the component neutron star (NS) masses ($M_1$, $M_2$), compactness ($C_1$, $C_2$), and ultimately the equation of state (EoS) of dense nuclear matter. The nuclear EoS describes the properties of dense matter, which is fundamental for understanding the behavior of NS throughout astrophysics.


The connection between the EoS parameters and ejecta properties is informed by numerical relativity (NR) Simulations. However, generating such simulations is computationally expensive and highly sensitive to the EoS adopted. Over the years, numerous studies have proposed distinct scaling relations to model key properties of BNS mergers, such as the accretion disk mass ($M_{\mathrm{disk}}$), the dynamical ejecta mass ($M_{\mathrm{dyn}}$), and the average ejecta velocity, as functions of underlying EOS parameters calibrated on existing NR simulation datasets.

Radice et al. (2018) ~\cite{Radice2018} first identified a correlation between the mass of the remnant accretion disk and the binary tidal deformability. Coughlin et al. (2019)~\cite{Coughlin2019} incorporated the threshold mass $M_{\mathrm{thr}}$, above which the remnant promptly collapses to a BH, while Dietrich et al. (2020)~\cite{Dietrich2020} extended this approach with a larger dataset, introducing explicit dependence on the mass ratio $q$. Krüger and Foucart (2020)~\cite{Kruger2020} highlighted the dominant role of the lighter star’s compactness $C_1$. Building on broader NR compilations, Nedora et al. (2021)~\cite{Nedora2021} proposed polynomial fits in $q$ and $\tilde{\Lambda}$. More recently, Lund et al. (2025) ~\cite{Lund2025} reaffirmed the strong $C_1$ dependence with an updated formula calibrated on the most extensive dataset to date \cite{Camilletti2024,Camilletti2022,Nedora2019,Nedora2021,Kruger2020}.

 Despite their widespread use, analytical fits for neutron star merger ejecta face serious limitations in reliability and generalization. Differences between models often exceed reported uncertainties, extrapolation beyond calibration can produce unphysical results, and uncertainty quantification remains limited \cite{Henkel2023}. These fits are also highly sensitive to the adopted microphysics and EOS \cite{Nedora2021}, and commonly used dynamical ejecta models fail for GW190425-like systems \cite{Camilletti2022}, underscoring the need for more robust approaches. In this work, we present symbolic regression (SR) algorithms \cite{Makke2024} that can autonomously discover analytical expressions that provide interpretability and better generalization across data sets.

\section{Symbolic Regression \& Numerical Relativity Simulation Dataset}



We focus our analysis specifically on the $M_{\mathrm{disk}}$, which represents a particularly challenging quantity to estimate due to its reported values from simulations \cite{Henkel2023} being highly sensitive to the time at which the measurement is performed, as well as to the criteria used to distinguish the accretion disk from the central remnant in neutron star–disk systems.

\textbf{Data:} We adopt two primary datasets of NR simulations. For training (calibration), we employ the same dataset utilized in \cite{Kruger2020} (hereafter Krüger20), which includes 56 NR simulations from \cite{Radice2018} and \cite{Kiuchi2019}. This dataset spans disk masses in the range $10^{-4}$--$0.234\,M_\odot$ and mass ratios $q = 0.77$--1.0. For evaluation, we use the NR simulation compilation from  \cite{Nedora2021} (hereafter Nedora21), which includes all simulations from the \cite{Kruger2020} dataset along with 63 additional simulations from multiple sources. This expanded test set covers a wider physical regime, with disk masses ranging from $3 \times 10^{-5}$--$0.30\,M_\odot$ and mass ratios from $0.55$--1.0. The analytical fitting formulae proposed by \texttt{Radice18} \cite{Radice2018}, \texttt{Coughlin19} \cite{Coughlin2019}, \texttt{Dietrich20} \cite{Dietrich2020}, \texttt{Krüger20} \cite{Kruger2020}, and \texttt{Lund25} \cite{Lund2025} are provided in Appendix \ref{app:fitting_formulae} \footnote{Throughout this paper, we label the two neutron stars with subscripts 1 and 2. The individual gravitational masses are denoted by $M_1$ and $M_2$, while the corresponding baryonic masses are written as $M_{b1}$ and $M_{b2}$. The total gravitational mass is defined as $M_{\mathrm{tot}} = M_1 + M_2$, and the mass ratio is given by $q = M_1 / M_2 < 1.0$. The dimensionless tidal deformability is denoted by $\tilde{\Lambda}$, and the compactnesses of the two stars are represented by $C_1$ and $C_2$.}. The \texttt{Nedora21} dataset provides only \( M_{1,2} \), \( C_{1,2} \), and \( \Lambda \); therefore, we were unable to apply this test set to some of the fitting formulas proposed in the literature. This limitation will be addressed in future work.

\textbf{Methods:} Symbolic regression is a subfield of machine learning and genetic programming (GP) that aims to infer interpretable, closed-form mathematical expressions directly from data \cite{Angelis2023,Koza1994, Schmidt2009}. Unlike traditional regression techniques, which optimize parameters within a fixed model structure, SR searches over both the space of model structures and parameter values. Learning models in the form of simple mathematical expressions offers much more than just potentially improved predictive power \cite{SyMANTICMuthyala2025} but it also enhances human interpretability. Most SR algorithms start with a population of random expressions iteratively refined through mutation, crossover, simplification, and constant optimization \cite{cranmer23}. Models are typically evaluated with mean squared error under multi-objective optimization, balancing accuracy and simplicity until convergence, producing a set of candidate expressions ranked by error and complexity, where each operator, variable, or constant contributes one unit to the total complexity by default.

\textbf{SR Training Setup:} We evaluated two symbolic regression frameworks for disk mass prediction. \texttt{PyOperon} \cite{OPERONBurlacu2020} is a Python wrapper to Operon, a popular SR method that uses genetic programming to explore a hypothesis space of possible symbolic expressions; we used the official implementation (\href{https://github.com/heal-research/pyoperon}{link}) with default hyperparameters and operators \texttt{\{+, -, *, /, sin, sqrt, log, tanh\}}. \texttt{PySR} \cite{cranmer23} is an evolutionary SR framework that balances accuracy and complexity via a multi-objective loss, with GPU and multi-threading support; we used the official implementation (\href{https://github.com/MilesCranmer/PySR}{link}). For \texttt{PySR}, we tested three operator sets: Core (\texttt{\{+, -, *, /; sqrt, log, exp, abs\}}), Extended (\texttt{\{+, -, *, /, max, min; sqrt, log, exp, square, cube, abs\}}), and All (\texttt{\{+, -, *, /, max, min, pow; sqrt, log, exp, square, cube, abs, sin, cos, tan, sinh, cosh, tanh\}}).

We also explored a physics-informed approach using predefined templates with the post-merger disk mass written as a linear combination of symbolic sub-expressions tied to variable groups: 
  
\begin{equation}
    M_{\mathrm{disk,pred}} = a \cdot f(C_1, M_1) + b \cdot g(C_2, M_2) + c \cdot h(q, \tilde{\Lambda}) + d,
\end{equation}

where $a,b,c,d$ are free coefficients and $f,g,h$ are symbolic functions discovered by SR. This expression allows us to isolate and quantify the influence of physical parameters on the remnant disk mass. We additionally queried \texttt{ChatGPT-4.0} (see Appendix~\ref{sec:chatgpt})
for an alternative template inspired by literature fits:  

\begin{equation}
    M_{\mathrm{disk,pred}} = a_1 f(q_0, C_1) + a_2 g(C_1, \tilde{\Lambda}) + a_3 h(M_1, M_2) + a_4 t(q_0) + a_5,
\end{equation}

where $a_{1\dots5}$ are scalar coefficients and $f,g,h,t$ are symbolic functions learned by SR. Their structure draws inspiration from previous fitting relations such as \texttt{Krüger20}, \texttt{Radice18}, and \texttt{Nedora21}. All models were calibrated on Krüger20 dataset for 3000 iterations with a maximum expression size of 20 (10 for the physics-informed template) and tree depth up to 5.


\section{Results}


\textbf{Model Selection:} For the \texttt{PySR} approach, models were selected along the Pareto front, which reflects the trade-off between predictive accuracy and complexity. From each training configuration, we chose a single representative model with complexity below $10$, prioritizing those whose Mean Squared Error (MSE) remained stable as complexity increased, indicating robustness to overfitting. The resulting equations are listed in Appendix~\ref{app:sr_formulas}, each labeled with an identifier encoding its configuration (e.g., \texttt{PySR\_template\_ext} refers to a model trained on Krüger20). In the template-based approach, we restricted the selection to models with complexity $\leq20$, since the physics-informed template itself carries a baseline complexity of $13$ ($17$ for the LLM-based template).


For the \texttt{PyOperon} framework, we selected the expression with a complexity of $9$, chosen for its lower structural complexity and exclusive dependence on the compactness of the lighter neutron star ($C_1$):

\begin{equation}
    M_{disk}^{PyOperon} = 0.118824 - (0.142985 \times \sin(\sin(40.896317 \times C_1)))
\end{equation}

\begin{table}
\centering
\caption{Quantitative performance comparison of symbolic regression models and literature fitting formulae for predicting post-merger disk mass. Metrics include MSE, MAE, coefficient of determination ($R^2$), and Bayesian Information Criterion (BIC), along with the model complexity (Comp.) and dependence on specific physical parameters. All SR models were calibrated on the \texttt{Krüger20} dataset, and all fitting formulae were evaluated on the \texttt{Nedora21} test set.}
\label{tab:model_performance}
\begin{tabular}{lrrrrrl}
\toprule
           Model &      MSE &      MAE &       $R^2$ &         BIC &  Comp. &   Dependence \\
\midrule
     PyOperon            & 0.002697 & 0.035435 & 0.611964 & -660.945373	 & 9  & $C_1$ \\
     PySR\_Free\_Ext       & 0.002751 & 0.035714 & 0.604268 & -658.608259 & 9  & $C_1$, $M_1$ \\
     PySR\_Free\_M12      & 0.002789 & 0.037436 & 0.598510 & -637.772776 & 13 & $M_1$, $M_2$, $\tilde{\Lambda}$ \\
     PySR\_Template\_Ext   & 0.002827 & 0.036365 & 0.593230 & -602.764208  & 20 &  $C_1$, $C_2$, $\tilde{\Lambda}$ \\
     Lund25              & 0.002910 & 0.036369 & 0.581383 & -637.580875 & 12 & $C_1$ \\
     PySR\_Free\_All       & 0.003029 & 0.036696 & 0.564164 & -651.900268 & 8  & $C_1$ \\
     PySR\_Template\_LLM   & 0.003156 & 0.037854 & 0.545986 & -589.732988  & 20 & $C_1$, $M_2$ \\
     PySR\_Template\_All   & 0.003322 & 0.037620 & 0.522114 & -593.150872 & 20 & $C_1$, $\tilde{\Lambda}$ \\
     PySR\_Free\_Core      & 0.003340 & 0.038944 & 0.520834 & -640.621426 & 8  & $C_1$ \\
     Krüger20               & 0.003665 & 0.039238 & 0.472704 & -595.777066 & 15 & $C_1$, $M_1$ \\
     PySR\_Template\_Core  & 0.005263 & 0.046414 & 0.242850 & -528.877037 & 20 &$M_1$, $M_2$, $q$, $\tilde{\Lambda}$ \\
     Radice18              & 0.006179 & 0.057595 & 0.111014 & -524.063305 & 12 & $\tilde{\Lambda}$ \\
\bottomrule
\end{tabular}
\end{table}


\begin{figure}[ht]
    \centering
    \begin{subfigure}{0.48\textwidth}
        \centering
        \includegraphics[width=\linewidth]{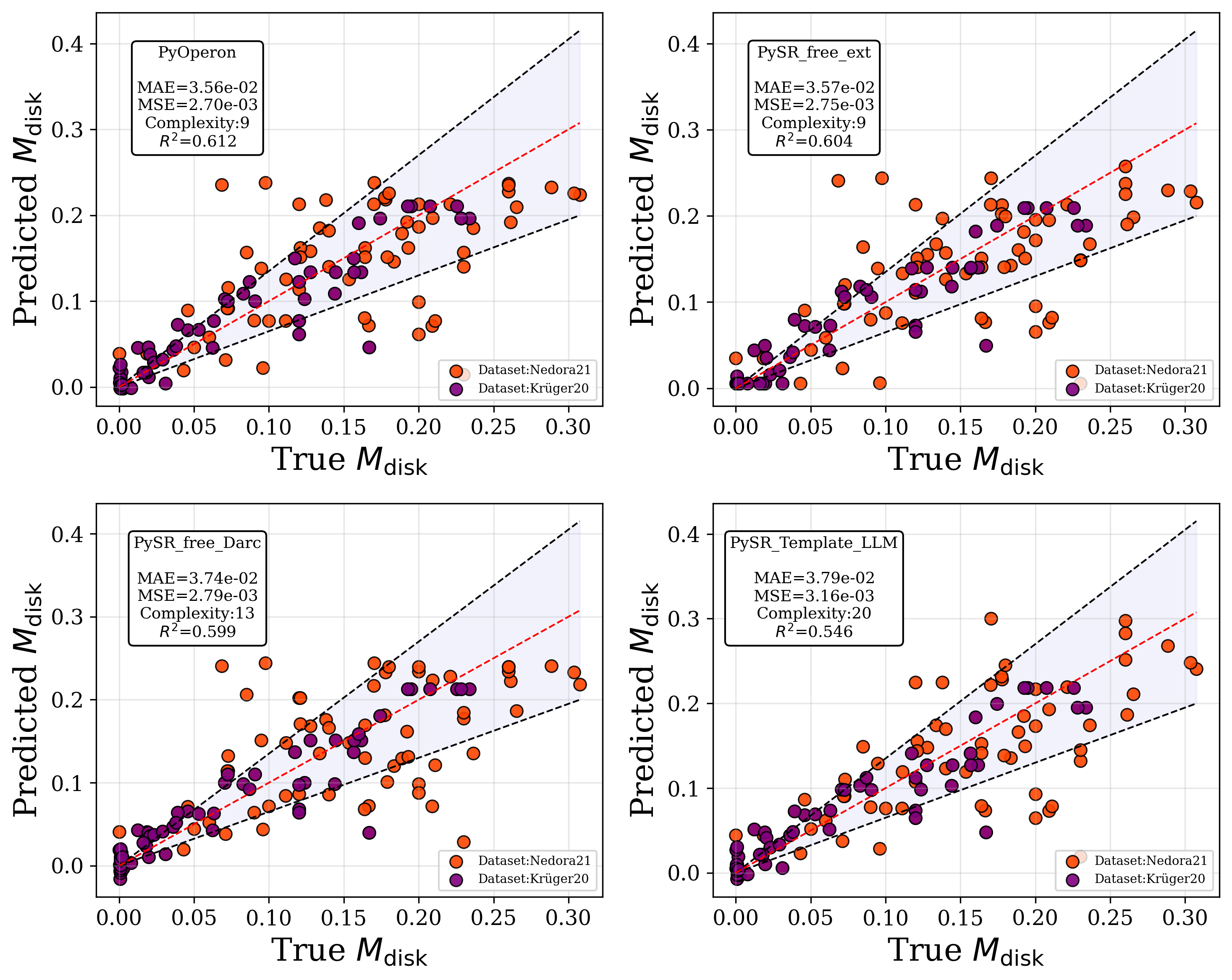}
        \caption{}
        \label{fig:sr_ytrue}
    \end{subfigure}
    \hfill
    \begin{subfigure}{0.48\textwidth}
        \centering
        \includegraphics[width=\linewidth]{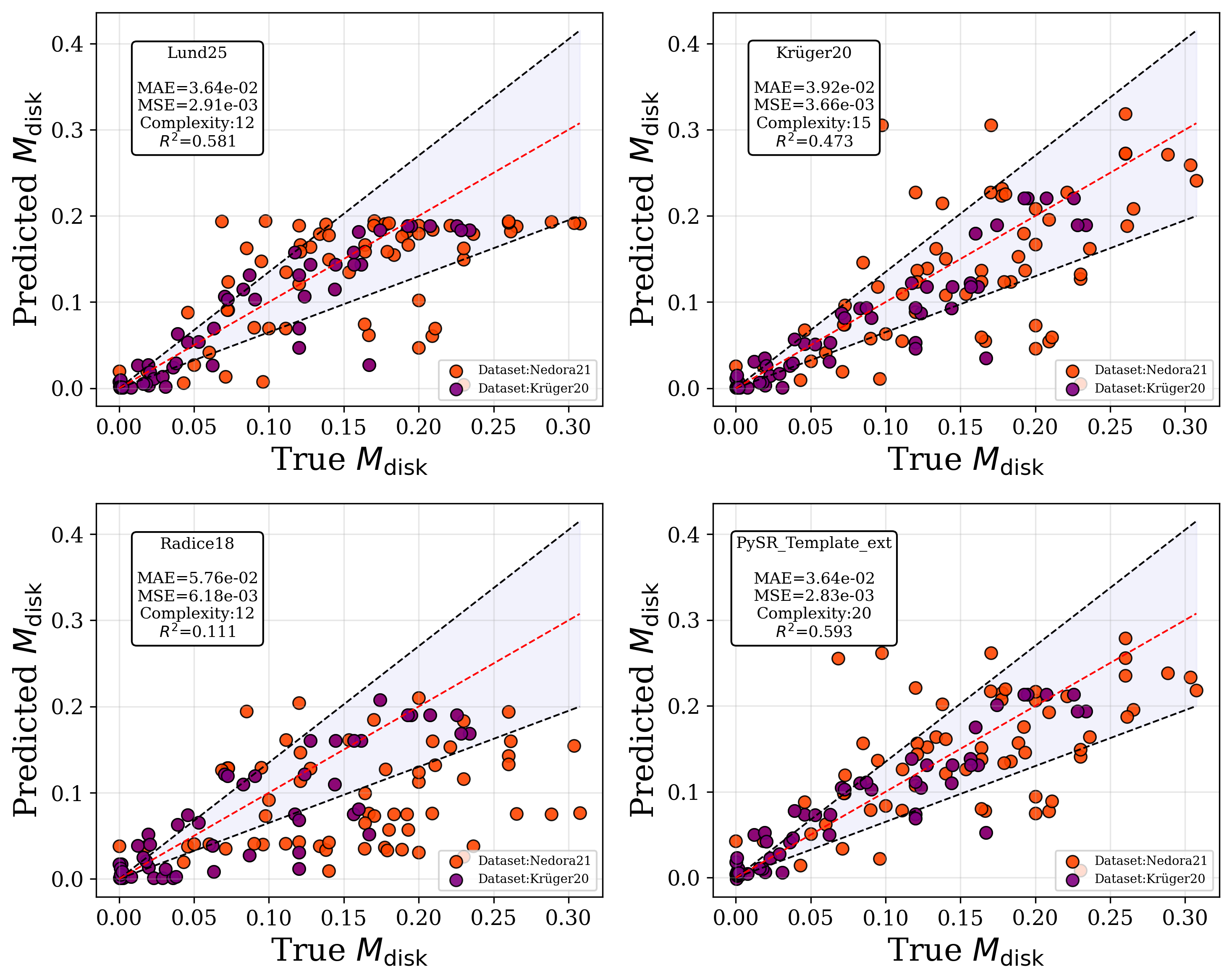}
        \caption{}
        \label{fig:lit_ytrue}
    \end{subfigure}
    
    \caption{Predicted disk mass versus disk mass from numerical relativity simulations. The left panel (\subref{fig:sr_ytrue}) presents results obtained with symbolic regression expressions, while the right panel (\subref{fig:lit_ytrue}) shows results from literature fitting formulae together with the \texttt{PySR} Template expression proposed by the LLM. Purple points correspond to the calibrated regime (Krüger20 dataset), whereas orange points represent the uncalibrated regime (Nedora21 dataset). Each inset box reports the model name, mean absolute error, mean squared error, expression complexity, and coefficient of determination ($R^2$).}
    \label{fig:combined_predxtrue}
\end{figure}


\textbf{Model Evaluation:} We compare literature fitting formulae with SR-derived expressions to evaluate whether SR models can (i) outperform existing fitting formulae in predictive accuracy, (ii) generalize beyond their calibration regime, (iii) recover physically meaningful dependencies, and (iv) produce more compact expressions while maintaining high accuracy. Model performance is assessed using mean absolute error (MAE), mean squared error (MSE), the Bayesian Information Criterion (BIC),
and the coefficient of determination ($R^2$). Table~\ref{tab:model_performance} summarizes the performance metrics, complexity, and parameter dependencies for all SR models trained on the \texttt{Krüger20} dataset, together with three representative literature formulae: \texttt{Krüger20}, \texttt{Radice18}, and \texttt{Lund25}. Nearly all SR-derived scaling relations outperform the \texttt{Krüger20} and \texttt{Radice18} fits on the \texttt{Nedora21} dataset, confirming that literature models often fail to generalize beyond their calibration regime \cite{Henkel2023}. Notably, four of the nine SR expressions also surpass \texttt{Lund25} across all metrics; in particular, \texttt{PyOperon} achieves a BIC more than 23 points lower, providing strong statistical evidence for the SR approach, even though it is evaluated on unseen data while \texttt{Lund25} is tested on its calibration set.

Although trained on the full set of available input variables ($M_{1,2}, C_{1,2}, \tilde{\Lambda}$), SR consistently rediscovers the dominant influence of the lighter star’s compactness ($C_1$), and in some cases also its mass ($M_1$), demonstrating both interpretability and robustness. Finally, Figures~\ref{fig:sr_ytrue} and~\ref{fig:lit_ytrue} compare predicted and true $M_{\mathrm{disk}}$, with purple markers denoting data beyond the calibration regime for the SR expressions. While \texttt{Lund25} is unable to reproduce cases with $M_{\mathrm{disk}} > 0.20\,M_\odot$,  models such as \texttt{PySR\_Free\_M12} and \texttt{PySR\_Template\_LLM} generalize effectively to this high-mass regime. In the intermediate disk-mass regime ($M_{\mathrm{disk}} \sim 0.05$--$0.20\,M_\odot$), the top SR models—\texttt{PyOperon} and \texttt{PySR\_Free\_Ext}—achieve MSE values of $2.12\times 10^{-3}$ and $2.09\times 10^{-3}$, outperforming \texttt{Krüger20} ($3.31\times 10^{-3}$) and matching \texttt{Lund25} ($1.95\times 10^{-3}$). This underscores SR’s ability to generalize beyond the training domain. Furthermore, SR can discover alternative predictor sets with comparable performance: e.g., \texttt{PySR\_Free\_M12}, which depends on $(M_1, M_2, \tilde{\Lambda})$, performs as well as \texttt{Lund25}, which relies solely on $C_1$. This flexibility enables the possibility of constraining previously inaccessible EOS parameters through disk mass measurements. 

In this work, we focused our analysis on performance metrics and model complexity. However, understanding the physical meaning of these models in extreme regimes is equally important. Since the existing fitting formulas are widely used in multi-messenger analyses, identifying new and reliable scaling relations requires exploring their qualitative behavior for large neutron star radii and high compactness, where numerical simulations are still limited. This aspect will be further investigated in the final version of this work. We suggest using the \texttt{PyOperon} model as an alternative to existing fitting formulas in the literature, given its strong performance and its dependence solely on the compactness parameter $C_1$.

\section{Conclusion}

Symbolic regression offers a promising framework for bridging data-driven modeling with scientific understanding. In this ongoing work, we evaluate two SR frameworks,\texttt{PySR} and \texttt{PyOperon}, using both physics-informed templates and free-form discovery to model post-merger disk mass in BNS mergers. Trained on a small dataset (56 NR simulations from \cite{Kruger2020}), the resulting expressions generalize effectively to the broader \cite{Nedora2021} compilation, demonstrating SR as an accurate and interpretable tool for uncovering physically meaningful relations in high-dimensional astrophysical data. Notably, models such as \texttt{PySR\_Free\_M12}, which depend on $M_1$, $M_2$, and $\tilde{\Lambda}$, achieve accuracy comparable to the literature while potentially enabling new parameter constraints from electromagnetic observations. 

In future work, we plan to extend this study by incorporating additional SR algorithms and methodologies, such as \texttt{AI-Feynman}\cite{AIFeynmanUdrescu2020} and \texttt{SciMED} \cite{SCIMEDKeren2023}. We also aim to calibrate these expressions on a more diverse and carefully curated dataset, including a broader range of NR simulations with more detailed physics. Another important direction involves applying SR to model the dynamical ejecta mass, with the goal of improving constraints on the neutron star equation of state from events such as GW170817 \cite{Nedora2021GW170817}.

\printbibliography

\appendix
\section*{Appendix A: Fitting Formulae from the Literature} \label{app:fitting_formulae}

We summarize below the main empirical models proposed in the literature to estimate the post-merger accretion disk mass $M_{\mathrm{disk}}$ as a function of binary and neutron star properties.

\paragraph{Radice et al. (2018) \cite{Radice2018}:}
\begin{equation}
M_{\mathrm{disk}} = \max\left(10^{-3}, \, \alpha + \beta \tanh\left( \frac{\tilde{\Lambda} - \gamma}{\delta} \right) \right),
\end{equation}
where $\alpha = 0.084$, $\beta = 0.127$, $\gamma = 567.1$, and $\delta = 405.14$.

\paragraph{Coughlin et al. (2019) \cite{Coughlin2019}:}
\begin{equation}
\log_{10} M_{\mathrm{disk}} = \max\left(-3, \, a \left(1 + b \tanh\left(\frac{c - M_{\mathrm{tot}}/M_{\mathrm{thr}}}{d} \right) \right) \right),
\end{equation}
where $a = -31.335$, $b = -0.9760$, $c = 1.0474$, and $d = 0.05957$.

\paragraph{Krüger \& Foucart (2020) \cite{Kruger2020}:}
\begin{equation}
M_{\mathrm{disk}} = M_1 \cdot \max\left(5 \times 10^{-4}, \left( a C_1 + c \right)^d \right),
\end{equation}
with $a = -8.1324$, $c = 1.4820$, and $d = 1.7784$.

\paragraph{Dietrich et al. (2020) \cite{Dietrich2020}:}
\begin{align}
q_{\mathrm{si}} &= \frac{1}{2} \tanh\left[ \beta(q_0 - Q_{\mathrm{trans}}) \right], \\
a &= a_0 + \Delta a \cdot q_{\mathrm{si}}, \qquad b = b_0 + \Delta b \cdot q_{\mathrm{si}}, \\
\log_{10} M_{\mathrm{disk}} &= \max\left(-3, \, a \left(1 + b \tanh\left( \frac{c - M_{\mathrm{tot}} / M_{\mathrm{thr}}}{d} \right) \right) \right),
\end{align}
where the coefficients are $a_0 = -1.581$, $\Delta a = -2.439$, $b_0 = -0.538$, $\Delta b = -0.406$, $c = 0.953$, $d = 0.0417$, $\beta = 3.910$, and $Q_{\mathrm{trans}} = 0.9$.

\paragraph{Lund et al. (2025) \cite{Lund2025}:}
\begin{equation}
\log_{10} M_{\mathrm{disk}} = \alpha \tanh(\beta C_1 + \gamma) + \delta,
\end{equation}
where $\alpha = -1.27$, $\beta = 68.01$, $\gamma = -11.72$, and $\delta = -1.98$.


\appendix
\section*{Appendix B: Symbolic Regression-Derived Expressions} \label{app:sr_formulas}

The following expressions were derived using symbolic regression calibrated on \cite{Kruger2020} dataset. 

\paragraph{PySR\_Free\_Core (Complexity 8):}
\begin{equation}
M_{\mathrm{disk}} = \left| \frac{0.1126}{C_1} - 0.6026 \right| - 0.0177
\end{equation}

\paragraph{PySR\_Free\_Ext (Complexity 9):}
\begin{equation}
M_{\mathrm{disk}} = \frac{\max \left[ M_1 (0.1750 - C_1), \, 0.00136 \right]}{0.2308}
\end{equation}

\paragraph{PySR\_Free\_All  (Complexity 8):}
\begin{equation}
M_{\mathrm{disk}} = \tanh\left( \max\left[ 0.00622, \frac{0.1763 - C_1}{0.1920} \right] \right)
\end{equation}

\paragraph{PySR\_Free\_M12 (Complexity 13):}
\begin{equation}
M_{\mathrm{disk}} = \min\left( 0.2676, \frac{0.0006162 \cdot \tilde{\Lambda}}{M_1^3} \right) - \frac{0.1508}{M_2^3}
\end{equation}


\paragraph{PySR\_Template\_LLM (Complexity 20, simplified to 16):}
\begin{align}
M_{\mathrm{disk}} \;=\;&
18170.047\, C_1^{6.911347}
\;-\; 6.031138\, e^{C_1} \nonumber \\
&-\; 0.02342195\, M_2
\;-\; 0.017214041 \times 432.73465
\;-\; 0.3142357 .
\end{align}

\section*{ChatGPT-4.0: Physics-Informed Symbolic Regression Prompt}\label{sec:chatgpt}
We fed the \cite{Henkel2023} study to the ChatGPT memory and used the following prompt:

\begin{verbatim}
[Start of Task – reset all prior context] 
You are a domain expert in relativistic astrophysics and gravitational 
wave modeling. Use only information derived from published, peer-reviewed 
physics literature related to binary neutron star (BNS) mergers and their 
associated ejecta (e.g., kilonovae, accretion disk formation). 

TASK: Based on the information from this article (and its cited references), 
synthesize a new physically motivated Ansatz — that is, a symbolic regression 
expression — for the accretion disk mass (M_{\mathrm{disk}}) formed after a BNS 
merger. This Ansatz will be used in PySR as a symbolic regression template 
function for fitting Numerical Relativity (NR) simulation outputs of BNS mergers. 
Your goal is to infer M_{\mathrm{disk}} using a closed-form expression based on 
physical parameters. 

Requirements:
Input Variables: Use a subset or full set of these physically meaningful input 
parameters:
    M_1: gravitational mass of neutron star 1 [solar masses]
    M_2: gravitational mass of neutron star 2 [solar masses]
    C_1: compactness of neutron star 1 [GM/(Rc^2)]
    C_2: compactness of neutron star 2
    q: mass ratio (M_2/M_1)
    \tilde{\Lambda}: reduced tidal deformability

Output Variable: 
    The expression should predict:
    M_{\mathrm{disk}}: accretion disk mass [solar masses]

Return Format: 
    Provide a complete PySR TemplateExpressionSpec object, e.g.,
    template = TemplateExpressionSpec(
        expressions=["f", "g", "h", "t"],
        variable_names=["M1", "M2", "C1", "C2", "q", "Lambda_tilde"],
        parameters={"p1": decide, "p2": decide},
        combine="template fitting function you created"
    )

Explanation Required:
    After providing the PySR template expression:
    - Justify the physical basis for the structure of each symbolic function
      (e.g., why q, \tilde{\Lambda}, and C are used together or multiplied).
    - cite examples or models from the literature 
      (e.g., Radice+2018, Dietrich+2017, Krüger+2020, Kiuchi+2019, etc.) 
      that support your formulation.
    - Discuss if the functional form is expected to extrapolate reasonably 
      outside the training domain, or whether it’s more phenomenological.

Your task is to derive a physics-informed symbolic expression for 
M_{\mathrm{disk}}, grounded in literature, with a structure usable 
in symbolic regression.
\end{verbatim}

ChatGPT-4.0 produced the following answer:

\begin{verbatim}
Based on the detailed analysis and comparative study presented in Henkel et al. (2022), 
and using insight from the fitting formulas in Krüger & Foucart (2020), Radice et al. (2018), 
Dietrich et al. (2020), and Nedora et al. (2021), I propose the following physics-informed 
Ansatz for the accretion disk mass M_{\mathrm{disk}} formed after a BNS merger.

Physical Motivation:
Accretion disk mass depends sensitively on:
    - The compactness C1 of the less massive neutron star — more compact stars 
      resist disruption.
    - The mass ratio q = M2 / M1, with lower values (more unequal masses) leading 
      to stronger tidal effects and larger disks.
    - The reduced tidal deformability \tilde{\Lambda}, which captures the binary’s 
      susceptibility to deformation.
    - A threshold behavior when approaching prompt collapse, typically captured 
      via a tanh transition (e.g., Dietrich+2020).

Proposed PySR Template:
template = TemplateExpressionSpec(
    expressions=["f", "g", "h", "t"],
    variable_names=["M1", "M2", "C1", "C2", "q", "Lambda_tilde"],
    parameters={"p1": 3, "p2": 2},
    combine="p1[1]*f(q, C1) + p1[2]*g(Lambda_tilde) 
             + p1[3]*h(M1, M2) + p2[1]*t(q) + p2[2]"
)
-----------------------------------------------------------------------------------------
\end{verbatim}

\end{document}